\documentclass[aps,prd,
preprintnumbers,nofootinbib,superscriptaddress]{revtex4}
\usepackage[dvips]{graphicx}
\usepackage{caption}
\usepackage{subcaption}
\usepackage{bm,latexsym,amsmath,amssymb,amsfonts,color}
\begin{document}

\title{Exact modulated hadronic tubes and layers at finite volume \\ in a cloud of $\pi$ and $\omega$ mesons}
\author{Gonzalo Barriga}
\email{gbarriga@cecs.cl}
\affiliation{Centro de Estudios Cient\'{\i}ficos (CECS), Casilla 1469, Valdivia, Chile}
\affiliation{Departamento de F\'{\i}sica, Universidad de Concepci\'{o}n,
Casilla 160-C, Concepci\'{o}n, Chile}
\author{Mat\'ias Torres}
\email{matiatorres@udec.cl}
\affiliation{Dipartimento di Fisica ``E. Pancini", Università di Napoli Federico II - INFN sezione di Napoli,
Complesso Universitario di Monte S. Angelo Edificio 6, via Cintia, 80126 Napoli}
\author{Aldo Vera}
\email{aldoveraseron@gmail.com}
\affiliation{Instituto de Ciencias Exactas y Naturales,Universidad Arturo Prat, Playa Brava 3256, 1111346, Iquique, Chile}
\affiliation{Facultad de Ciencias, Universidad Arturo Prat, Avenida Arturo Prat Chac\'on 2120, 1110939, Iquique, Chile}

\begin{abstract}
We construct topological soliton solutions describing baryonic tubes and layers with modulation in the $SU(2)$ non-linear sigma model coupled with $\omega$-mesons in $3+1$ dimensions. Using appropriate Asäntze for the pionic matter field and the $\omega$-mesons vector potential, the complete set of seven coupled partial differential equations can be solved analytically. These solutions represent modulated tubes and layers at finite volume with arbitrary baryon number, where the modulation of the solitons in one direction is determined by one of the three degrees of freedom of the pionic field, satisfying the equation of a two-dimensional free massless chiral scalar field.
As expected, the inclusion of the $\omega$-mesons to the Non-linear sigma model allows to reduce the repulsion energy between baryons, which leads to a flattening of the tubes and layers in one direction, forming a kind of ``nuclear linguine phase''. Also, we show that this construction can be carried out even when higher order terms in the large $N_c$ expansion are included -in particular the Skyrme term- without spoiling the integrability of the field equations.
\end{abstract}

\maketitle

\newpage

\section{Introduction}
Quantum Chromodynamics (QCD) constitutes one of the pillars of the standard model of particle physics. Featuring three colors and several flavors, QCD dynamics become strongly coupled at low energies while weak at high energies. It is in the strong sector where the non-perturbative nature of QCD remains inaccessible to standard analytical techniques, being the numerical methods the most used \cite{Shifman1}, \cite{Wilczek}. In this context, the development of innovative analytical techniques is of crucial importance, allowing a controlled study of the strong dynamics. One of the most important models to address the strong sector of QCD is known as the non-linear sigma model (NLSM), which manifests spontaneous chiral symmetry breaking and provides an accurate description of pions at low energies \cite{MantonBook}. 

Additionally, the NLSM admits the existence of topological soliton solutions, which are interpreted as baryons \cite{Lizzi}, \cite{Shifman2}, \cite{Witten0}, \cite{ANW}, with the topological charge equal to the baryonic number. In this context, baryons emerge from the non-linear interactions between mesons. However, these configurations are not energetically stable due to Derrick's theorem \cite{Derrick}. This problem can be solved including higher derivative corrections in the Lagrangian that come from the large $N_c$ expansion, being the Skyrme term the simplest one \cite{Skyrme} (see also \cite{MantonBook}, \cite{Zahed}). Nevertheless, it is important to highlight that stable topological solitons describing baryons can be constructed even without including the Skyrme term in the NLSM. In fact, this can be carried out by circumventing Derrick's theorem in different ways. For example, by working in a finite space or coupling the theory to spin-1 matter fields.\footnote{Here we consider two ways to avoid Derrick's theorem. First, we consider a system without spherical symmetry. In particular, we construct solitons confined to a finite volume in regular patterns. Second, the matter field depends on a light-like degree of freedom, which constitutes one of the main ingredients in the construction of analytical solutions.}

On the other hand, although the NLSM and Skyrme models provide a good description of hadrons at low energies, both in their static properties and as interacting states \cite{ANW}, \cite{Jackson1}, \cite{Ordonez}, \cite{Yan}, some predictions differ significantly from the experimental data. One of these discrepancies is the nuclear binding energy. The predictions from the Skyrme model point to a repulsion energy between baryons greater than what has been measured in experiments.

Now, is it possible to stabilize the solitons and, at the same time, reduce the expected value of the binding energy coming from the NLSM by introducing vector mesons to the theory. In fact, in Refs. \cite{Adkins1}, \cite{Jackson2}, \cite{Meissner} (see also \cite{[8]p}, \cite{[9]p}, \cite{[10]p}, \cite{[11]p}, \cite{[12]p} and references therein) has been shown that the inclusion of $\omega$-mesons accomplishes this task. In this paper, we construct analytical solutions in the NLSM coupled to $\omega$-mesons with non-trivial topological charge. These solutions represent ordered arrays of baryonic tubes and layers at finite volume. Then, we generalize our results to the case where the Skyrme term is included, and even when higher order corrections in the t' Hooft expansion are considered \cite{tHooft}, that is, the generalized Skyrme model \cite{Witten1}, \cite{Gudnason}, \cite{Scherer}, \cite{Adam1} (see also \cite{Marleau1}, \cite{Marleau2}, \cite{Marleau3}).

It is well known that, when baryonic matter is under extreme conditions, ordered arrays are expected to appear as a result of the non-linear interactions between the constituents. This has been explored using numerical methods in the Skyrme model in Refs. \cite{Klebanov}, \cite{Goldhaber}, \cite{Kugler1}, \cite{Braaten}, \cite{Kugler2}, \cite{Adam2}, \cite{Adam3}. Recently, in Refs. \cite{crystal1}, \cite{crystal2}, \cite{crystal3}, \cite{crystal4}, \cite{crystal5}, \cite{Alvarez}, \cite{Hidalgo} (see also \cite{Barriga1}, \cite{SU(N)1}, \cite{SU(N)2}, \cite{Torres}), the first analytical solutions describing crystals of topological solitons at finite volume in the NLSM and Skyrme models were constructed (see \cite{Rebolledo} for a review). One of the main achievements of such construction is that the configurations obtained are very similar to the nuclear pasta phases \cite{Dorso} (see also \cite{pasta1}, \cite{pasta2}, \cite{pasta2a}, \cite{pasta2b}, 
\cite{pasta3}, \cite{pasta4}, \cite{pasta5}, \cite{pasta6}, \cite{pasta7}, 
\cite{pasta8}, \cite{pasta9} and references therein), in particular, nuclear lasagna (in the form of layers) and nuclear spaghetti (in the form of tubes). This is of particular interest as nuclear pasta states are expected to emerge by subjecting baryonic matter to extreme conditions, for example, in supernovae cores and in the crust of neutron stars, where densities exceed the normal nuclear density \cite{Watanabe}, \cite{Liebling}.\footnote{In addition to the Skyrme model, in the study of compact stars, the Walecka model is a very useful theory \cite{Walecka1}, \cite{Walecka2}, \cite{Walecka3}, which also describes nucleons and mesons. A discussion about the relation between these models can be found in Ref. \cite{Barriga2}.}

Although the study of the formation of nuclear pasta phases has been approached using simulations such as molecular dynamics \cite{Watanabe} and numerical methods (see, for instance, \cite{aprox0}, \cite%
{aprox1}, \cite{aprox2}, \cite{aprox3}, \cite{aprox4}, \cite{aprox5}, \cite%
{aprox6}, \cite{aprox7}, \cite{aprox8}, \cite{aprox9}, \cite{aprox10}), until now there has been no analytical approach to this problem, which would open an important window in the understanding of baryonic matter at extreme conditions. In this work, we aim in that direction. We generalize the solutions constructed in Ref. \cite{Barriga2}, showing that baryonic tubes and layers made of $\pi$ and $\omega$ mesons can be promoted to solutions of the generalized Skyrme model. The inclusion of an arbitrary light-like degree of freedom in the matter field allows for modulation of the solitons in one direction, and the time evolution of these configurations can be shown explicitly.

The paper is organized as follows: In Section II, we introduce the NLSM coupled to $\omega$-mesons. In Section III, we construct analytical solutions describing baryonic tubes and layers at finite volume, and discuss their main physical properties. In Section IV, we show how the inclusion of the $\omega$-mesons allows for a reduction in the binding energy between baryons. Also, we discuss the differences and similarities between our solutions and the crystals of gauged skyrmions.  In Section V, we show that the inclusion of higher-order corrections to the theory does not spoil the integrability of the equations. Section VI is devoted to the conclusions.

\section{The model}

The NLSM coupled to vector $\omega$-mesons is described by the action
\begin{gather} \label{I}
I[U, \omega]=\int d^{4} x \sqrt{-g}\left(\frac{K}{4} \operatorname{Tr}\left[L^{\mu} L_{\mu}\right] -\frac{1}{4} S_{\mu \nu} S^{\mu \nu}-\frac{1}{2} M_{\omega}^{2} \omega_{\mu} \omega^{\mu}-\gamma \rho_{\mu} \omega^{\mu}\right) \ , \\
L_{\mu}=U^{-1} \nabla_{\mu} U= L_{\mu}^{j} t_{j} \ , \quad S_{\mu\nu}= \nabla_\mu \omega_\nu - \nabla_\nu \omega_\mu \ , \quad t_{j}=i \sigma_{j} \ ,  \notag
\end{gather}
where $U(x) \in SU(2)$ is the pionic field, $\sigma_i$ are the Pauli matrices, $\omega_{\mu}$ is the 4th-vector potential describing the $\omega$-mesons, $\nabla_{\mu}$ is the covariant derivative, $K$ and $\gamma$ are positive coupling constants fixed experimentally, and $M_{\omega}$ corresponds to the $\omega$-mesons mass. In our convention $c=\hbar=1$,  Greek indices run over the four dimensional space-time with a mostly plus signature, and Latin indices are reserved for those of the internal space. 

The field equations of the system are a set of seven coupled partial differential equations given as follows: First, the three equations obtained varying the actions with respect to the pionic matter field $U$ are 
\begin{equation}
      \nabla_{\mu}L^{\mu}-\frac{6 \gamma}{K}\nabla_{\nu}(\epsilon^{\mu\nu\lambda\rho}\omega_{\mu}L_{\lambda}L_{\rho})=0 \ . \label{NLSM}
\end{equation}
Second, the field equations that come from the coupling with the $\omega$-mesons, obtained through the variation with respect to the field $\omega_\mu$, are
\begin{equation}
\nabla_{\mu} S^{\mu \nu}-M_{\omega}^{2} \omega^{\nu}=\gamma \rho^{\nu} \ .\label{omegaeq}
\end{equation}
The $\omega$-mesons interact with the $\pi$-mesons through the topological current, $\rho^{\mu}$, present in Eq. \eqref{I}, which is defined as
\begin{align}
\rho^{\mu}=\epsilon^{\mu\nu\lambda\rho} \operatorname{Tr}\left[\left(U^{-1} \partial_{\nu} U\right)\left(U^{-1} \partial_{\lambda} U\right)\left(U^{-1} \partial_{\rho} U\right)\right .] \ . 
\label{topcurrent}
\end{align}
where $\epsilon^{\mu\nu\lambda\rho}$ is the Levi-Civita tensor.
The integral over a space-like hypersurface $\Sigma$ of the $\rho^0$ component of the topological current leads to the topological charge
\begin{align}
B=\frac{1}{24 \pi^{2}} \int_{\Sigma} \rho^{0} \ , \label{B}
\end{align} 
which determines the baryonic number of a given matter configuration.

On the other hand, the energy-momentum tensor of the theory is given by
\begin{equation}
\begin{aligned}
T_{\mu\nu}=-\frac{K}{2} & \operatorname{Tr}\left[L_\mu L_\nu-\frac{1}{2} g_{\mu \nu} L^\alpha L_\alpha\right]+S_{\mu \alpha} S_\nu{ }^\alpha-\frac{1}{4} S_{\alpha \beta} S^{\alpha \beta} g_{\mu \nu} \\
& +M_\omega^2\left(\omega_\mu \omega_\nu-\frac{1}{2} g_{\mu \nu} \omega^\alpha \omega_\alpha\right)+\gamma\left(\rho_\mu \omega_\nu+\rho_\nu \omega_\mu-g_{\mu \nu} \rho_\alpha \omega^\alpha\right) \  .
\end{aligned}
\end{equation}

\section{The solutions}

In this section, we construct two types of analytical solutions of the NLSM coupled to $\omega$-mesons using two different Ans\"atze for the pionic field $U(x)\in SU(2)$. The first one, which describes baryonic layers (the so-called ``lasagna phase"),  is built via the Euler angles representation, while the second one, which describes baryonic tubes (the so-called ``spaghetti phase"), is constructed via the exponential representation. Both configurations possess a non-vanishing topological charge.\footnote{
As we will see below, the inclusion of the $\omega$-mesons 
induces a change in the geometry of the nuclear pasta states; in particular, the baryonic tubes flatten in one direction. This is the reason for calling these novel solutions ``nuclear linguine phase".}
For the $\omega$-mesons field, we will use a convenient choice that allows decoupling the degrees of freedom corresponding to each type of meson. This desirable characteristic is achieved by Ans\"atze that satisfy the following relation: 
\begin{equation}
   \nabla_{\nu}(\epsilon^{\mu\nu\lambda\rho}\omega_{\mu}L_{\lambda}L_{\rho})=0  \label{desacoplar}\  , 
\end{equation}
which is clear from Eq. \eqref{NLSM}. We will see that, although this condition may at first seem restrictive, the solutions that emerge from it exhibit particular characteristics that come from the coupling of the $\pi$-mesons and $\omega$-mesons.

As we are interested in the construction of analytical solutions at finite volume, we will consider the metric of a box
\begin{equation} \label{Box}
    ds^2= - dt^2 + L_x^2 dx^2 + L_y^2 dy^2 + L_z^2 dz^2 \ , 
\end{equation}
where the adimensional spatial coordinates $\{x,y,z\}$ have the ranges 
\begin{equation} \label{ranges}
0 \leq x \leq 2\pi \ ,  \quad 0 \leq y \leq \pi \ , \quad 0 \leq z \leq 2\pi \ ,
\end{equation}
and the coefficients $L_i$ fix the size of the box in which the solitons are confined.

\subsection{Modulated baryonic layers in a cloud of $\pi$ and $\omega$ mesons}

For the construction of analytical solutions describing baryonic layers at finite volume in the NLSM coupled to $\omega$-mesons, we will use an Ansatz inspired in the case of Yang-Mills theory and the Skyrme model introduced in Refs. \cite{Alvarez}, \cite{Hidalgo}, \cite{SU(N)1}, \cite{SU(N)2} (see also \cite{Aviles}, \cite{Oh}, \cite{Ayon1}, \cite{Ayon2}, \cite{Flores}, \cite{Pais}). In those references, it has been shown that the parameterization in Euler angles is convenient to describe finite volume configurations homogeneous in two spatial dimensions. 

An element of $SU(2)$ can be written in the Euler angles representation as follows 
\begin{equation}
    U=e^{F(x^\mu) t_{3}}e^{H(x^\mu) t_{2}}e^{G(x^\mu)t_{3}} \label{lasagnaansatz} \ , 
\end{equation}
where $F(x^\mu)$, $G(x^\mu)$, $H(x^\mu)$ are the three degrees of freedom of the pionic field, and they are, in principle, arbitrary functions of the coordinates. A choice for the degrees of freedom that allows to reduce the NLSM equations to a decoupled system at finite volume is the following (see \cite{Hidalgo} for details)  
\begin{equation}
    F(x^{\mu})= q y \ ,  \quad H(x^{\mu})=H(x) \ , \quad G(x^{\mu})=G(t,z) \ . \label{layers}
\end{equation}
In fact, the above Ansatz, in absence of the $\omega$-mesons field, reduces the NLSM equations to the following solvable system  
\begin{gather}
\partial_x^2 H=0 \ , \label{EqH} \\
\Box G= \partial_t^2 G - \frac{1}{L_z^2} \partial_z^2 G = 0 \ .  \label{EqG1} 
\end{gather}
Here below we point out some important comments about these equations.
First, being Eq. \eqref{EqH} a simple ODE and Eq. \eqref{EqG1} the wave equation, the solutions of these equations are, respectively 
\begin{equation}  \label{sol1}
H(x)=\frac{(1+2n)}{4}x \ , 
\end{equation}
where $n$ is an integer number fixed by the boundary conditions, and 
\begin{equation}
     G=G_{-}+G_{+} \ , \label{G}
\end{equation}
where
\begin{align*}
G_{+}&=z_0^{+}+v_{+}\left(\frac{t}{L_z}+z\right)+\sum_{n \neq 0}\left(a_n^{+} \sin \left[n\left(\frac{t}{L_z}+z\right)\right]+b_n^{+} \cos \left[n\left(\frac{t}{L_z}+z\right)\right]\right) \ , \\ G_{-}&=z_0^{-}+v_{-}\left(\frac{t}{L_z}-z\right)+\sum_{n \neq 0}\left(a_n^{-} \sin \left[n\left(\frac{t}{L_z}-z\right)\right]+b_n^{-} \cos \left[n\left(\frac{t}{L_z}-z\right)\right]\right) \ .
\end{align*}
Now, once the interaction with the $\omega$-field is taken into account, in order to decouple the contribution of the $\omega$-mesons from the pionic degrees of freedom, we have to impose
\begin{gather}
    \left(\partial_t G\right)^2-\frac{1}{L_z^2}\left(\partial_z G\right)^2=\left(\partial_t G+\frac{1}{L_z} \partial_z G\right)\left(\partial_t G-\frac{1}{L_z} \partial_z G\right)=0 \ .  \label{chiraleq}
\end{gather}
The above condition along with the following Ansatz for the $\omega$-mesons 
\begin{equation}
    \omega_\mu=-\frac{u}{p}\left(\partial_{z}G, 0,0,L_{z}^2\partial_{t}G\right),\quad u=u(x) \ , \label{omega1}
\end{equation}
(with $p$ an integer number) guarantees that the full system of equations remains integrable. In fact, the potential in Eq. \eqref{omega1} satisfies the constraint in Eq. \eqref{desacoplar}.   It must be highlighted that Eq. \eqref{chiraleq} is not inconsistent with the wave equation in Eq. \eqref{EqG1}. Instead of that, Eq. \eqref{chiraleq} is a particular case of Eq. \eqref{EqG1}; it projects one of the modes $G_{-}$ or $G_{+}$ to zero. Therefore, the main difference with respect to Ref. \cite{Hidalgo} is that the degree of freedom $G$ now describes a two-dimensional free massless chiral scalar field.

Note that this vector field is very similar to the Ansatz introduced in Refs. \cite{crystal2} and \cite{crystal3} for the Maxwell potential. However, there are relevant differences between these two cases. We will discuss this point in the next section.
From the above, the four equations related to the $\omega$-mesons are reduced to just one differential equation, namely
\begin{equation}  
   u''-L_x^2 M_{\omega}^2 u =\frac{12 \gamma p q L_x }{L_y L_z} \sin (2 H) \partial_x H  \ . \label{Equ}
\end{equation}
This equation can be easily solved due to the profile $H$ is a linear function. In fact, the $\omega$-mesons profile turns out to be
\begin{equation}
    u(x)= -\frac{12 \gamma L_{x} (2 n+1) p q }{L_{y} L_{z} \left(4 L_{x}^4 M_{\omega}^2+(2 n+1)^2\right)} \sin \left(\left(n+\frac{1}{2}\right) x\right) \ .
\end{equation}
where we have used periodic boundary conditions for the $u$ function; $u(0)=u(2\pi)=0$. The boundary conditions for the fields $H$ and $G$ come from imposing the topological charge in Eq. \eqref{B} will be an integer number. In fact, using the previous parametrization in Eqs. \eqref{lasagnaansatz} and \eqref{layers}, the topological charge density $\rho^{0}$ becomes 
\begin{equation*}
    \rho^{0}=\frac{12 q }{L_{x}L_{y}L_{z}} \sin (2 H) \partial_x H \partial_z G \ .
\end{equation*}
It follows that, choosing the following boundary conditions
\begin{equation}
H(2\pi)=\frac{(1+2n)}{2} \pi\ , \quad H(0)=0 \ , \qquad G(t, z=2\pi)-G(t, z= 0)=(2 \pi) p \ ,  \label{BC1}
\end{equation}
the topological charge turns out to be $B= p q$. Therefore, the baryon number for these kinds of solutions can be an arbitrary integer number.
Although the integer parameter $n$ does not contribute to the topological charge, it determines the number of layers in the lattice along the $x$ direction. 

Note that the boundary conditions in Eq. \eqref{BC1} implies that the coefficients $v_{\pm}$ in Eq. \eqref{G} satisfy $ p=v_{+}-v_{-}$, while the coefficients $\{a^{\pm},b^{\pm}\}$ do not contribute to the topological charge.

\begin{figure}[ht!]
     \centering
     \begin{subfigure}[h!]{0.3\textwidth}
         \centering \includegraphics[width=1\textwidth]{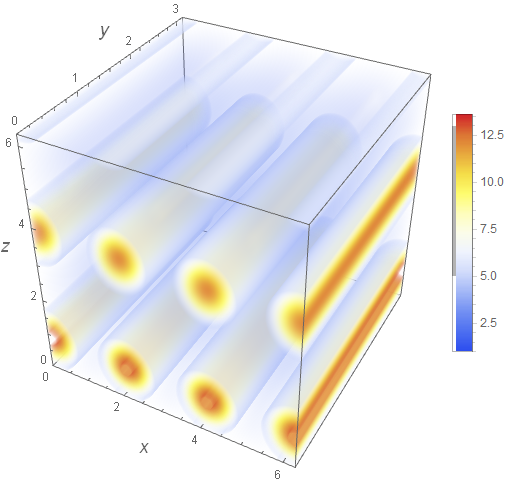}
         \label{fig:lasagna1}
     \end{subfigure}
     \qquad
     \begin{subfigure}[h!]{0.3\textwidth}
         \centering  \includegraphics[width=1.\textwidth]{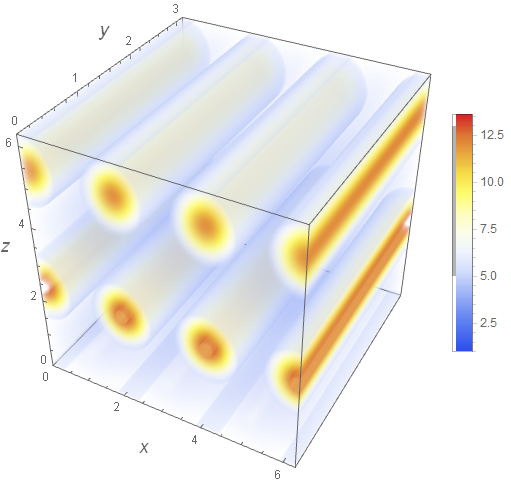}
         \label{fig:lasagna2}
     \end{subfigure}
     
       \vskip\baselineskip
       \centering
       
     \begin{subfigure}[h!]{0.3\textwidth}
         \centering \includegraphics[width=1.\textwidth]{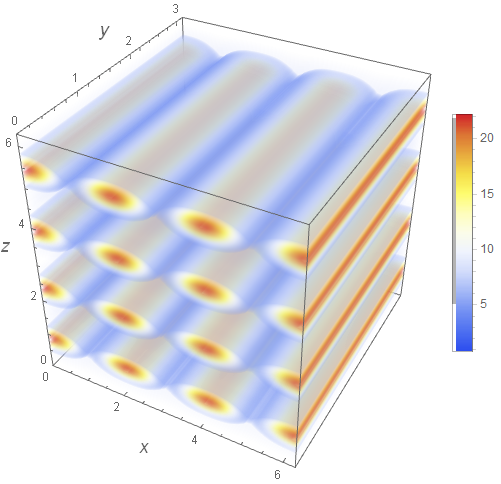}
         \label{fig:lasagna3}
     \end{subfigure}
     \qquad
     \begin{subfigure}[h!]{0.3\textwidth}
         \centering  \includegraphics[width=1.\textwidth]{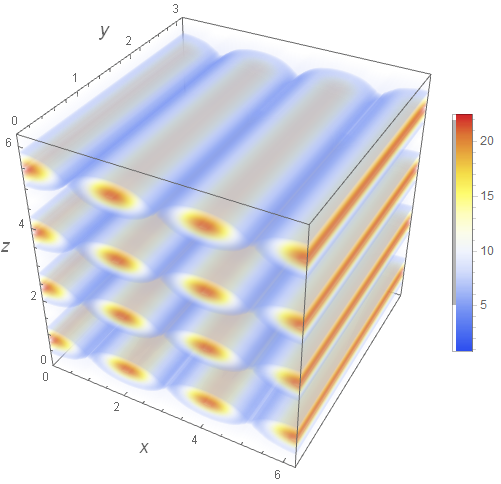}
         \label{fig:lasagna4}
     \end{subfigure}
   \caption{Energy density of baryonic layers with 
   $q=n=1$ and $p=2$  In the upper row we have chosen the modulation coefficients as  $b_{1}=0.1$, $b_{2}=0.5$, while the lower row is set to $a_{1}=0.1$, $a_{2}=0.5$. The left column corresponds to a snapshot at $t=0$ while the right column is set at time $t=\frac{3\pi}{2}$. We have assumed $K=1$ and $L_x=L_{y}=L_z=1$.}
        \label{fig:lasagnas}   
\end{figure}

The energy density, $\varepsilon=T_{00}$, for the solutions presented above is given by
\begin{align}
   \varepsilon= \frac{K}{2}  \left(\frac{H'^2}{L_{x}^2}+\frac{q^2}{L_{y}^2}+\Delta G \right)+\frac{L_{z}^2}{2L_{x}^2p^2} u'^2\Delta G+\frac{1}{2p^2} M_{\omega}^2 u^2 L_{z}^2\Delta G+\frac{12L_{z} \gamma  q   }{L_{x}L_{y}p}u H' \sin (2 H)\Delta G \ , \label{ED1}
\end{align}
where $\Delta G =\partial_{t}G^2+\frac{1}{L_{z}^2}\partial_{z}G^2$, is the energy density of a free massless scalar field, and the prime denotes $\partial_x$. Fig. \ref{fig:lasagnas} shows the energy density of these configurations. One can see that the solution describes an array of baryonic layers at finite volume; the number of layers is determined by the number $n$ in the boundary conditions, while the number of baryons is fixed by $p$ and $q$. The modulation of the tubes is controlled by the modes associated with the $G$ function in Eq. \eqref{G}, as well as the evolution in time. 

\subsection{Modulated baryonic tubes in a cloud of $\pi$ and $\omega$ mesons} 

Crystals of baryonic tubes as solutions of the NLSM and Skyrme model can be constructed using the exponential representation, as have been shown in Refs. \cite{crystal1}, \cite{crystal2}, \cite{crystal3} and \cite{crystal4} (see also \cite{Hidalgo}, \cite{SU(N)1}, \cite{SU(N)2}).

An element of $SU(2)$ in the exponential representation is written as
\begin{gather}
U^{\pm 1}(x^{\mu })=\cos \left( \alpha \right) \mathbf{1}_{2}\pm \sin \left(
\alpha \right) n^{i}t_{i}\ ,\ \ \ \ n^{i}n_{i}=1\ ,  \label{hedgehog} \\
n^{1}=\sin \Theta \cos \Phi \ ,\ \ \ n^{2}=\sin \Theta \sin \Phi \ ,\ \ \
n^{3}=\cos \Theta \ ,  \notag
\end{gather}%
where $\alpha$, $\Theta$ and $\Phi$ are the three degrees of freedom of $SU(2)$. Following Refs. \cite{crystal2} and \cite{Hidalgo}, we will choose these functions as
\begin{equation}
\alpha =\alpha (x)\ ,\qquad \Theta = Q y\ ,\qquad \Phi =G(t,z) \ . \label{tubesansatz}
\end{equation}
As in the case of the baryonic layers presented above, one can check that there is an Ansatz that satisfies the constraint in Eq. \eqref{desacoplar} and, therefore, allows to decouple the NLSM equations from the $\omega$-mesons equations; that is, 
\begin{equation}
    \omega_\mu=-\frac{v}{p}\left(\partial_{z}G, 0,0,L_{z}^2\partial_{t}G\right),\quad v=v(x,y) \ , \label{omega2}
\end{equation}
with $p$ a constant. This potential is very similar to Eq. \eqref{omega1}, but in this case the function $v$ must depends on two spatial coordinates instead of just one.

Replacing Eqs. \eqref{Box}, \eqref{tubesansatz} and \eqref{omega2} into the NLSM equations, we obtain the following decoupled system
\begin{gather} \label{Eqalpha}
    \alpha'' -Q^2 \frac{L_x^2}{L_y^2}\sin(\alpha) \cos(\alpha) = 0 \ , \\
    \Box G = \partial_t^2 G - \frac{1}{L_z^2} \partial_z^2 G = 0 \ , \label{BoxG}  \\
  (\partial_t G)^2-\frac{1}{L_z^2} (\partial_z G) ^2 = \biggl(\partial_t G + \frac{1}{L_z}\partial_z G\biggl)\biggl(\partial_t G - \frac{1}{L_z}\partial_z G\biggl) = 0 \ . \label{EqG0}
\end{gather}
Again, Eqs. \eqref{BoxG} and \eqref{EqG0} are solved by one of the modes expansion in Eq. \eqref{G}, defining a free massless chiral field theory in $1+1$ dimensions for the $G$ field. The simplest solution of this system is a linear function $G=\frac{t}{L_z}-z$, which has been explored in Ref. \cite{Barriga2}; these are tubes without modulation. On the other hand, Eq. \eqref{Eqalpha} can be solved analytically in terms of Elliptic functions. 
Even more, the explicit solution of this equation is not necessary since all the relevant quantities that characterize the solution (such as the energy density and the topological charge density) only depend on $\alpha$ and its derivatives, not on the $x$ coordinate explicitly. Indeed, Eq. \eqref{Eqalpha} can be reduced to the following quadrature: 
\begin{equation}
\alpha'^2+ \frac{L_x^2 Q^2 }{2 L_y^2}\cos (2 \alpha )=E_0 \ , 
\end{equation}
where one can read $\alpha'$ in terms of $\alpha$ (here $E_0$ is an integration constant fixed by the boundary conditions).
On the other hand, the $\omega$-mesons equations are reduced to the following partial differential equation
\begin{gather} \label{Eqv}
    (-\Delta + M_\omega^2) v = f(x,y) \ ,  \\ 
    f(x,y)=  \frac{12 \gamma Qp}{L_x L_y L_z} \sin^2(\alpha)\sin(Qy) \alpha' \ , \qquad \Delta = \frac{1}{L_x^2}\partial_x^2 +\frac{1}{L_y^2}\partial_y^2 \ .  \notag
\end{gather}

Although this equation is not as simple as that of the nuclear lasagna phase, Eq. \eqref{Eqv} is a Poisson equation, and its general solution can be written as
\begin{equation}
v(\vec{r})=\int d \vec{r}^{\prime} \mathrm{G}\left(\vec{r}, \vec{r}^{\prime}\right) f\left(\vec{r}^{\prime}\right) \ , \qquad \left(-\nabla_{\vec{r}}^2+M_\omega^2\right) \mathrm{G}\left(\vec{r}, \vec{r}^{\prime}\right)=\delta\left(\vec{r}-\vec{r}^{\prime}\right) \ ,
\end{equation}
where $\vec{r}=(x,y)$ and $\mathrm{G}\left(\vec{r}, \vec{r}^{\prime}\right)$ is the corresponding Green function. 
The topological charge density of this configuration is given by 
\begin{equation*}
    \rho^0=-\frac{12 Q}{L_x L_y L_z} \sin (Q y) \sin ^2(\alpha) \alpha^{\prime} \partial_z G  \ .
\end{equation*}
In order to have an integer value for the baryon number, we must impose the following boundary conditions:
\begin{align}
    \alpha(2 \pi)-\alpha(0)=n \pi, \quad G(t, z=0)-G(t, z=2 \pi)=(2 \pi) p \ ,
\end{align}
so that, the baryon number for these configurations turns out to be $B=np$.  Note that the parameter $Q$ in Eq. \eqref{tubesansatz} must be an odd number to ensure that the baryon number does not vanishes.

The respective energy density reads 
\begin{equation}
    \begin{aligned}
            \varepsilon=&\frac{K}{2} \left(\frac{\sin ^2(\alpha ) \left(\Delta G L_y^2 \sin ^2(Q y  )+Q^2\right)}{L_y^2}+\frac{2 \alpha '^2}{L_x^2}\right) +  \frac{\Delta G L_z^2 \left(L_y^2( \partial_x v)^2+L_x^2 (\partial_y v)^2\right)}{2 L_y^2 L_x^2 p^2}\\
            & +\frac{\Delta G L_z^2 M_\omega^2 v^2}{2 p^2}  -  \frac{12 L_z \gamma Q \sin (Q y) \alpha ' \sin ^2(\alpha ) v  \Delta G}{ L_x L_y  p} \ , 
    \end{aligned}
\end{equation}
where $\Delta G $ has been defined below Eq. \eqref{ED1}.
\begin{figure}[ht!]
     \centering
     \begin{subfigure}[h!]{0.3\textwidth}
         \centering \includegraphics[width=1\textwidth]{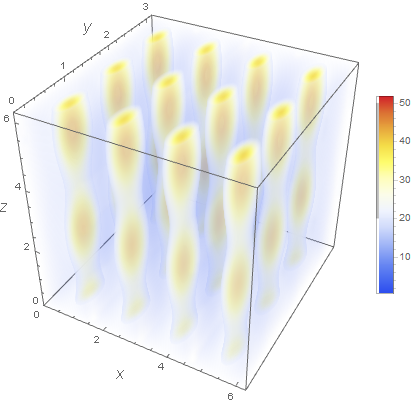}
         \label{fig:linguini1}
     \end{subfigure}
     \qquad
     \begin{subfigure}[h!]{0.3\textwidth}
         \centering  \includegraphics[width=1\textwidth]{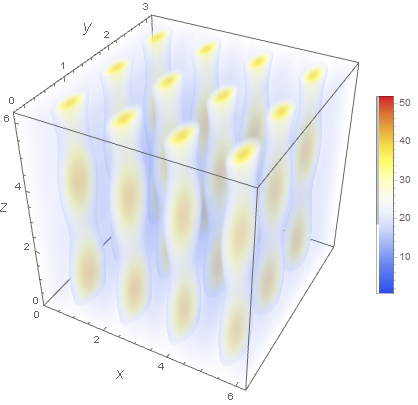}
         \label{fig:linguini2}
     \end{subfigure}
       \vskip\baselineskip
       \centering
     \begin{subfigure}[h!]{0.3\textwidth}
         \centering \includegraphics[width=1\textwidth]{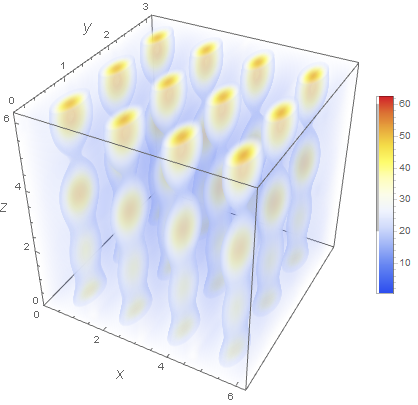}
         \label{fig:linguini3}
     \end{subfigure}
     \qquad
     \begin{subfigure}[h!]{0.3\textwidth}
         \centering  \includegraphics[width=1\textwidth]{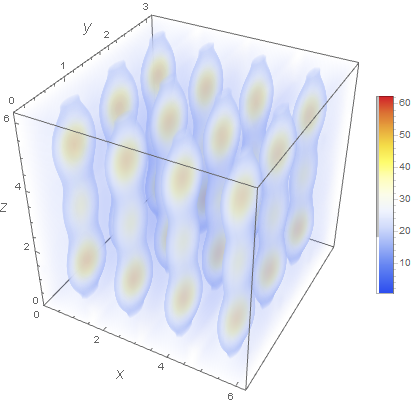}
         \label{fig:linguini4}
     \end{subfigure}
        \caption{Energy density of baryonic tubes with 
   $n=4$, $Q=3$ and $p=6$. In the upper row we have chosen the modulation coefficients as $b_{1}=0.1$, $b_{2}=0.5$, while the lower row is set to $a_{1}=0.1$, $a_{2}=0.5$. The left column corresponds to a snapshot at $t=0$ while the right column is set at time $t=\frac{3\pi}{2}$. We have assumed $K=1$ and $L_x=L_{y}=L_z=1$.}
        \label{fig:linguinis}
\end{figure}
Fig. \ref{fig:linguinis} shows the energy density of these configurations. One can see that the system describes a lattice of baryonic tubes, where the numbers $n$ and $p$ define the baryon number, that is, the number of tubes in the $x$ direction. On the other hand, the parameter $Q$ repeats the pattern in the $y$ direction. 

An interesting issue arises from the coupling with the $\omega$-mesons. It is expected that nuclear spaghetti-like solutions are tubes extended in the $z$ direction (with or without modulation) whose cross sections are concentric circles \cite{Dorso}, \cite{Watanabe}. In fact, this is what is expected from nuclear spaghetti phases, as can be seen from the simulations obtained in Refs. \cite{pasta1}, \cite{pasta2}, \cite{pasta2a}, \cite{pasta2b}, 
\cite{pasta3}, \cite{pasta4}, \cite{pasta5}, \cite{pasta6}, \cite{pasta7}, 
\cite{pasta8}, \cite{pasta9}. Here, however, as the configurations extend in the $z$ direction, the cross sections no longer remain spherical but take on oval shapes. This phenomenon appears precisely because the coupling of the $\omega$-mesons reduces the repulsion energy between the baryons. Indeed, in Fig. \ref{fig:linguinis} we can see that, for a fixed baryon number (the number of tubes in the $x$ direction) the tubes in the $y$ direction move closer together due to the lower repulsion which generates the $\omega$-mesons in addition to the $\pi$-mesons.

\section{Comparing crystals}

As was proposed in Refs. \cite{Adkins1}, \cite{Jackson2}, \cite{Meissner}, the inclusion of vector mesons in the NLSM allows for a reduction in the predicted binding energy for nucleons, which makes it more compatible with the experimental data. Even more, this can be clearly seen from the analytical solutions presented in the previous section, where in the case of the baryonic tubes (see Fig. \ref{fig:linguinis}), a flattening emerges due to the presence of the $\omega$-mesons.
\begin{figure}
\begin{subfigure}[h!]{0.45\textwidth}
         \centering \includegraphics[width=1\textwidth]{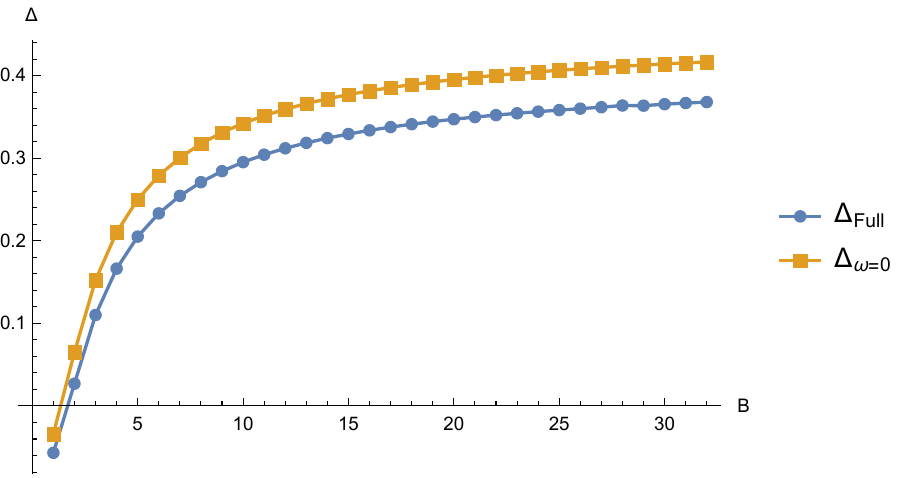}
\end{subfigure}
\begin{subfigure}[h!]{0.4\textwidth}
         \centering \includegraphics[width=1\textwidth]{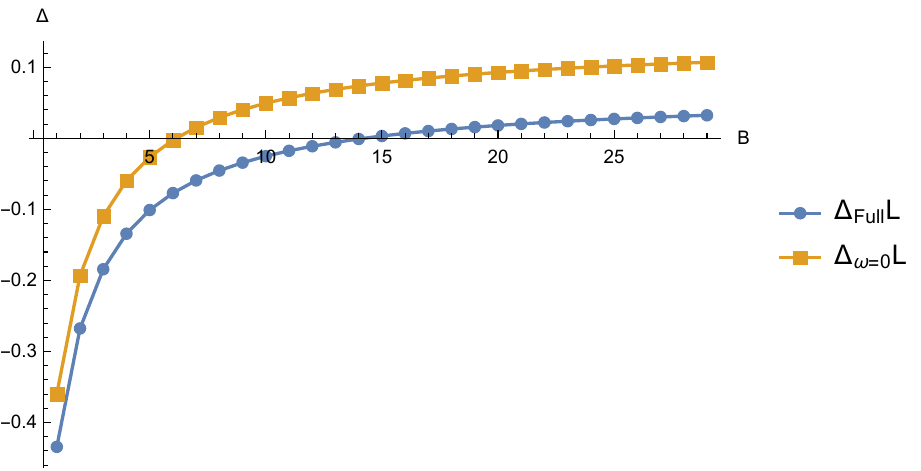}
\end{subfigure}\\
\begin{subfigure}{0.45\textwidth}
    \includegraphics[width=1\textwidth]{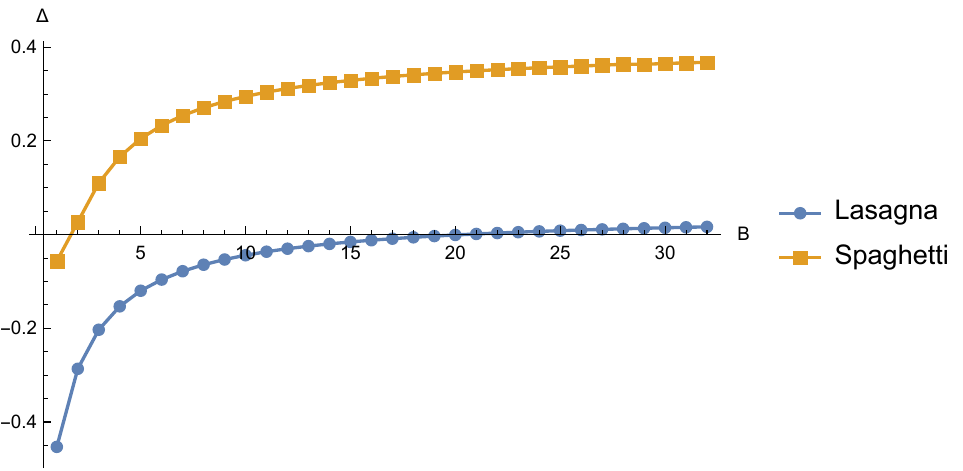}
\end{subfigure}
\caption{ $\Delta(B)$ for the tubes (upper left) and the layers (upper right) in both cases: The NLSM  coupled with the $\omega$-mesons $\left(\Delta_{\text {Full }}\right)$ and the NLSM  without the $\omega$-mesons $\left(\Delta_{\omega=0}\right)$. The comparison between lasagna and spaghetti phases (below). Here, we have set $K=2$, $\gamma=0.1$, $p=q=1$, $L_x=L_y=L_z=1$ and $M_\omega=0$.}
\label{fig:comparison}
\end{figure}

Another way to see it is by introducing the quantity $\Delta(B)$, which measures the interaction energy between baryons \cite{Adkins1}. This quantity is defined as
\begin{equation}
    \Delta(B) = \frac{E_{(B+1)}-(E_{(B)}+E_{(1)})}{(B+1) E_{(1)}} \ , 
\end{equation}
where $B$, in our cases, is the baryon number for the baryonic tubes and layers and $E_{(i)}$ is the total energy of the system containing $(i)$ baryons. This quantity is an increasing function of $B$, due to the strong short-range repulsion between baryons. In fact, from Fig. \ref{fig:comparison} one can see that the inclusion of the $\omega$-mesons reduces the interaction energy since these curves associated with the solutions with $\omega$-mesons are below the one that only contains $\pi$-mesons.

Another interesting fact comes from the comparison between the baryonic tubes and layers. From Fig. \ref{fig:comparison} (below), one can see that, when comparing the $\Delta(B)$ function for both configurations for a fixed baryon number, the layer pattern is the one that most reduces the repulsion energy between the baryons that constitute the system. Tube-like configurations are more repulsive, at least in this sector.

At this point, it is important to highlight the difference between the baryonic crystals coupled to $\omega$-mesons presented here versus the gauged crystals shown in Refs. \cite{crystal2}, \cite{Hidalgo}, constructed using similar methods. First, let us remember that gauged skyrmions come from the minimal coupling of the pionic field with photons through the covariant derivative, which is defined as  
\begin{equation*}
D_{\mu }U=\nabla _{\mu }U+A_{\mu }U\hat{O}\ ,\quad \hat{O}=U^{-1}\left[
t_{3},U\right] \ .
\end{equation*}
An appropriate Ansatz for the Maxwell potential allows to decouple the Skyrme equations from the Maxwell equations (just as in the case of the $\omega$-mesons that we have shown here), allowing the construction of crystalline structures of baryonic tubes and layers (see \cite{crystal2}, \cite{crystal3}, \cite{crystal4}, \cite{Aviles}, \cite{Oh}). However, the fact that the coupling with vector mesons comes from an interaction term in the action instead of the minimal coupling, implies the following relevant differences: 
\begin{enumerate}
\item For crystals of gauged skyrmions, the Skyrme equations are affected by the electromagnetic coupling (even when the three equations reduce to a single equation for the skyrmion profile), while in the case of the crystals with $\omega$-mesons the equation for the profile is exactly the same with and without $\omega$-mesons. This can be seen in Refs. \cite{SU(N)1}, \cite{SU(N)2}, \cite{Aviles}, \cite{Oh} for the nuclear lasagna phase.
\item In the case of gauged skyrmions, the Maxwell's equations reduce to a Schr\"odinger-like equation, while for the $\omega$-mesons, the equations are reduced to a Poisson equation. 
This can be seen in Refs. \cite{crystal2}, \cite{crystal3}, \cite{crystal4}, \cite{SU(N)1}, \cite{SU(N)2} for the nuclear spaghetti phase.
\item While the coupling with the $\omega$-mesons generates a flattering in the energy density of the baryonic tubes and layers, the coupling with the electromagnetic field changes the intensity of the energy inside the tubes or layers, but the geometry remains the same.
\end{enumerate}

\section{Higher order corrections}

In this section, we will show that the previous set of solutions can be constructed even when higher-order derivative terms are included in the action. For this purpose, we consider the generalized Skyrme model \cite{tHooft}, \cite{Witten1}, \cite{Gudnason}, \cite{Scherer}, \cite{Adam1} (this is, the NLSM model plus the Skyrme term and higher order corrections) coupled to $\omega$-mesons, described by the action  
\begin{equation}
I_{\text{gen}}[U, \omega]=\int d^4 x \sqrt{-g}\left[K \operatorname{Tr}\left(L^\mu L_\mu+\frac{\lambda}{8} F_{\mu \nu} F^{\mu \nu}\right)-S_{\mu \nu} S^{\mu \nu}-\frac{1}{2} M_{\omega}^{2} \omega_{\mu} \omega^{\mu}-\gamma \rho_{\mu} \omega^{\mu}+\mathcal{L}_{\text {corr }}\right] \ , \label{skyrme corrections}
\end{equation} 
where $F_{\mu \nu}=\left[L_\mu, L_\nu\right]$. 
The term $\mathcal{L}_{\text {corr }}$ represents the subleading corrections to the Skyrme model, which can be obtained via chiral perturbation theory (see \cite{Scherer} and references therein) or by the large $N_c$ expansion of QCD \cite{tHooft}, \cite{Witten1}. To make the calculations clearer, we will consider only the first correction, which is
\begin{equation} \label{L6}
\mathcal{L}_{\text {corr }} =\frac{c_6}{96} \operatorname{Tr}\left[F_\mu{ }^\nu F_\nu{ }^\rho F_\rho{ }^\mu\right] \ , 
\end{equation}
(with $c_6$ a coupling constant) although our results are still valid even including the next higher order terms.

The field equations obtained by varying the action in Eq. \eqref{skyrme corrections} with respect to the $U$ field, are 
\begin{equation}
\nabla^\mu L_\mu+\frac{\lambda}{4} \nabla^\mu\left[L^\nu, F_{\mu \nu}\right]-\frac{6\gamma}{K}  \nabla_\nu\left(\epsilon^{\mu \nu \lambda \rho} \omega_\mu L_\lambda L_\rho\right)+\frac{6 c_6}{K}\left[L_\mu, \partial_\nu\left[F^{\rho \nu}, F_\rho{ }^\mu\right]\right] =0 \ ,
\label{fieldeqscorrected}
\end{equation}
 while the $\omega$-mesons equations are the same as in Eq. \eqref{omegaeq}.

First, using the same Ansatz for the baryonic layers defined in Eqs. \eqref{lasagnaansatz} and \eqref{layers}, one can check that the coupled system in Eq. \eqref{fieldeqscorrected} reduces to the same relations in Eqs. \eqref{EqH}, \eqref{EqG1} and \eqref{chiraleq}. The contribution that comes from the Skyrme term is encoded in a global factor, namely  
\begin{equation}
    K \left(L_y^2 -\lambda  p^2\right) \partial_x^2 H=0 \ .
\end{equation}
Note that the contribution in Eq. \eqref{L6} does not appear at all in the field equations. This is because the correction in Eq. \eqref{L6} is the topological charge density to the square, and it vanishes when (at least) one of the pionic degrees of freedom is light-like, as is indeed the case for our solutions. 

Something similar happens with the baryonic tubes.
Indeed, for the Ansatz in Eqs. \eqref{hedgehog} and \eqref{tubesansatz}, the system in Eq. \eqref{fieldeqscorrected}  is reduced to the same relations in Eqs. \eqref{BoxG} and \eqref{EqG0} for the $G$ function, together with a single  ODE for the  profile $\alpha$:
\begin{equation}
   \alpha '' \left(\lambda  Q^2 \sin ^2(\alpha )-L_y^2\right)+\lambda  Q^2  \sin (\alpha ) \cos (\alpha )\alpha '^2+L_x^2 Q^2 \sin (\alpha) \cos (\alpha )=0 \ .
\end{equation}
This last equation can be solved in terms of generalized Elliptic Integrals, but again, it can lead to a first order ODE:
\begin{gather}
\partial_x\left(Y(\alpha)\left(\alpha^{\prime}\right)^2+W(\alpha)+E_0\right)=0 \ ,  \\
Y(\alpha)=2 L_y^2-\lambda  Q^2+\lambda  Q^2 \cos (2 \alpha ), \quad W(\alpha)= L_x^2 Q^2 \cos (2 \alpha ) \ , \notag
\end{gather}
(where $E_0$ is an integration constant fixed by the boundary conditions), allowing all the relevant quantities to be written only in terms of $\alpha$, and not explicitly on the coordinates.

\section{Conclusions}

We have shown that exact solutions describing crystals of baryonic tubes and layers can be constructed in a NLSM that includes $\pi$-mesons and $\omega$-mesons. These configurations have an arbitrary topological charge (written as the product of two integer numbers) and can be modulated through a light-like degree of freedom present in both mesonic fields. The inclusion of the $\omega$-mesons to the NLSM allows for a reduction in the binding energy between baryons, making the predictions of the model more compatible with the experimental data.  Interestingly, with the solutions constructed here, this can be seen very clearly: the energy density plot of the tube-like solutions shows that the tubes flatten in one direction by reducing the repulsion between baryons and forming a sort of ``linguine phase". Also, by looking at the same plot, it could be possible to  break the tubes into small pieces and form a ``gnocchi phase" by choosing a fine-tuning of the mode's coefficients. In this way, the inclusion of the $\omega$-mesons could be important for an analytical description of the gnocchi phase.
Finally, we have shown that these exact solutions can be constructed in the generalized Skyrme model, where higher-order derivative terms are included in the action.

\acknowledgments

G.B. is funded by the National Agency for Research and Development ANID grant 21222098. 
M. T. is funded by Agencia Nacional de Investigación y Desarrollo (ANID) grant 72210390.
A.V. has been partially funded by National Agency for Research and Development ANID SIA SA77210097.
A.V thanks the hospitality of ICEN (UNAP) where part of this project was done.

\end{document}